\documentclass[pdflatex,sn-mathphys-num]{sn-jnl}

\usepackage{graphicx}%
\usepackage{multirow}%
\usepackage{amsmath,amssymb,amsfonts}%
\usepackage{amsthm}%
\usepackage{mathrsfs}%
\usepackage[title]{appendix}%
\usepackage{xcolor}%
\usepackage{textcomp}%
\usepackage{manyfoot}%
\usepackage{booktabs}%
\usepackage{algorithm}%
\usepackage{algpseudocode}%
\usepackage{listings}%

\usepackage{soul}%
\usepackage{subcaption}%
\usepackage{hyperref}%
\usepackage{url}%
\usepackage{makecell}%

\usepackage{comment}%

\theoremstyle{thmstyleone}%
\theoremstyle{thmstyletwo}%

\theoremstyle{thmstylethree}%

\begin{document}

\title[Article Title]{Ethic-BERT: An Enhanced Deep Learning Model for Ethical and Non-Ethical Content Classification}


\author[]{\fnm{Mahamodul Hasan} \sur{Mahadi}}\email{24-93530-3@student.aiub.edu}
\author[]{\fnm{Md. Nasif} \sur{Safwan}}\email{22-49041-3@student.aiub.edu}
\author[]{\fnm{Souhardo} \sur{Rahman}}\email{22-49068-3@student.aiub.edu}
\author[]{\fnm{Shahnaj} \sur{Parvin}}\email{sparvin@aiub.edu}
\author[]{\fnm{Aminun} \sur{Nahar}}\email{aminun.nahar@aiub.edu}
\author*[]{\fnm{Kamruddin} \sur{Nur}}\email{kamruddin@aiub.edu}

\affil[]{\orgdiv{Department of Computer Science}, \orgname{American International University-Bangladesh}, \orgaddress{\city{Dhaka}, \postcode{1229}, \country{Bangladesh}}}


\abstract{Developing AI systems capable of nuanced ethical reasoning is critical as they increasingly influence human decisions, yet existing models often rely on superficial correlations rather than principled moral understanding. This paper introduces Ethic-BERT, a BERT-based model for ethical content classification across four domains: Commonsense, Justice, Virtue, and Deontology. Leveraging the ETHICS dataset, our approach integrates robust preprocessing to address vocabulary sparsity and contextual ambiguities, alongside advanced fine-tuning strategies like full model unfreezing, gradient accumulation, and adaptive learning rate scheduling. To evaluate robustness, we employ an adversarially filtered 'Hard Test' split, isolating complex ethical dilemmas. Experimental results demonstrate Ethic-BERT’s superiority over baseline models, achieving 82.32\% average accuracy on the standard test, with notable improvements in Justice and Virtue. In addition, the proposed Ethic-BERT attains 15.28\% average accuracy improvement in the HardTest. These findings contribute to performance improvement and reliable decision-making using bias-aware preprocessing and proposed enhanced AI model.
}




\maketitle

\section{Introduction}
\label{sec:introduction}

Incorporating ethical reasoning into artificial intelligence systems is a vital step toward creating technology that is both responsible and aligned with societal values. As AI systems play an increasingly prominent role in mediating human interactions and making autonomous decisions, it is crucial to ensure they operate within the bounds of ethical principles \cite{woodgate2024macro, sholla2024ethical}. However, encoding the complexity of human moral reasoning into AI systems poses significant challenges, requiring innovative approaches to bridge the gap between human values and computational frameworks \cite{li2022ethics}.

Ethical morality involves the principles of right and wrong that guide human behavior, encompassing dimensions such as justice, fairness, well-being, duties, and virtues. These principles are deeply interconnected, often leading to conflicts that require nuanced decision-making. Humans rely on cultural, social, and personal contexts to navigate moral ambiguities, but replicating this capacity in AI systems demands sophisticated techniques \cite{akbarmuhammad, han2022aligning}. The integration of ethical reasoning into AI is particularly important because of its potential societal impact \cite{hauer2022importance}. AI systems, if left unchecked, can amplify biases, produce harmful outputs, or make decisions that conflict with shared human values \cite{hendrycks2021ethics}. To address these issues, researchers have turned to text-based scenarios as a means of evaluating AI systems' ability to understand and apply ethical reasoning. These text based scenarios allows for the representation of complex moral dilemmas, providing a practical medium for assessing how AI aligns with human ethical judgment.

Recent advancements in NLP, particularly the development of transformer architectures, have made it possible to achieve significant progress in understanding context and intent in textual data. Datasets like ETHICS \cite{dataset} leverage these advancements, presenting scenarios derived from philosophical theories, including justice, deontology, virtue ethics, utilitarianism, and commonsense morality. These benchmarks challenge AI systems to move beyond simple pattern recognition and address the moral complexity inherent in real-world scenarios. Despite these advancements, progress in embedding ethical reasoning into AI has been limited \cite{vaswani2017attention}. Models trained on ETHICS dataset \cite{dataset} have shown some promise but struggle with nuanced scenarios and adversarial examples. 

The key challenges of achieving better results in ethical reasoning tasks include:
\begin{itemize}
    \item Lack of high-quality datasets that reduce ambiguity and enhance representativeness.
    \item Existing models struggle with nuanced ethical reasoning, limiting accuracy in moral decision-making.
    \item AI models rely on spurious correlations rather than deep moral reasoning, leading to misclassifications in complex ethical scenarios.
    \item The dataset primarily reflects Western moral perspectives, reducing its applicability to diverse cultural and ethical viewpoints.
\end{itemize}

In this research, we address these key challenges by the following key contributions:
\begin{enumerate}
    \item Advanced model architectures utilizing state of the art transformer models and fine-tuning techniques to strengthen ethical reasoning capabilities.
    \item Comprehensive evaluation demonstrating significant performance gains on standard and adversarially filtered test sets.
    \item Enhanced dataset preparation to address ambiguities and introduce diverse contextual scenarios, improving data quality.
\end{enumerate}

By advancing the interplay between ethical reasoning and AI, our work lays the groundwork for systems that are more aligned with human values and equipped to handle the complexities of real-world moral dilemmas.

\section{Related Work}
The increasing volume of digital content has created a pressing need for effective ethical content classification to manage misinformation, hate speech, and inappropriate material. Recent research efforts have focused on developing robust detection techniques using machine learning (ML) and deep learning (DL) models to improve accuracy, efficiency, and adaptability. At the same time, the ethical and moral implications of content have also become crucial, requiring models capable of analyzing not only spam content but also ethically unacceptable text. This review explores significant contributions in this field, focusing on advancements in detecting and mitigating harmful content while preserving contextual integrity.

In the domain of explicit content detection, Bhatti et al. \cite{bhatti2018} proposed an Explicit Content Detection (ECD) system targeting NSFW content using a residual network-based deep learning model. Their approach, integrating YCbCr color space and skin tone detection, achieved 95\% accuracy in classifying explicit images and videos. Similarly, Khandekar et al. \cite{khandekar2023} focused on NLP techniques for detecting unethical and offensive text, leveraging LSTM and BiLSTM networks, which outperformed traditional models with an accuracy of 86.4\%. Horne et al. \cite{horne2023} discussed the ethical challenges in automated fake news detection, emphasizing algorithmic bias and lack of generalizability. Their analysis of 381,000 news articles revealed the limitations of detection models that overfit benchmark datasets. Kiritchenko and Nejadgholi \cite{kiritchenko2020} introduced an “Ethics by Design” framework for abusive content detection, highlighting fairness, explainability, and bias mitigation. Their two-step process categorized identity-related content before assessing severity, reinforcing the importance of ethical considerations in content moderation. Schramowski et al. \cite{schramowski2022} examined the moral biases embedded in large pre-trained models like BERT, demonstrating how these biases could be leveraged to steer text generation away from toxicity. They identified a “moral direction” within the embedding space, which could be used to rate the normativity of text. This aligns with the ethical text assessment aspect of our research. Mittal et al. \cite{mittal2023} conducted a comparative study on deep learning models for hate speech detection, concluding that fine-tuned RoBERTa models outperformed CNNs and BiLSTMs in a ternary classification system. Mnassri et al. \cite{mnassri2023} explored a multi-task learning framework that integrated emotional features with hate speech detection using BERT and mBERT. Their approach improved performance by leveraging shared representations across tasks, reducing overfitting and false positives. Saleh et al. \cite{saleh2023} investigated the effectiveness of domain-specific word embeddings in hate speech detection, concluding that while specialized embeddings enhanced detection of coded hate speech, pre-trained BERT models achieved the highest F1-score with 96\%. Jim et al. \cite{JIM2024100059} review advancements in sentiment analysis, highlighting machine learning, deep learning, and large language models. They explore applications, datasets, challenges, and future research directions to enhance performance.

Sultan et al. \cite{sultan2023} analyzed shallow and deep learning techniques for cyberbullying detection across social media platforms. Their study, comparing six shallow learning algorithms with three deep models, found that BiLSTM models achieved the best recall and accuracy. This underscores the challenges in identifying masked or subtle offensive content, emphasizing the need for sophisticated models. Wadud et al. \cite{csse.2023.027841} examine methods for offensive text classification, emphasizing the need for improved multilingual detection. They introduce Deep-BERT, a model combining CNN and BERT, which enhances accuracy in identifying offensive content across different languages. Also, spam detection has been widely explored using traditional ML models and DL approaches. Similarly, Maqsood et al. \cite{maqsood2023} proposed a hybrid approach combining Random Forest, Multinomial Naive Bayes, and SVM with CNNs, observing that SVM outperformed other traditional ML models, while CNNs excelled on larger datasets. Guo et al. \cite{guo2022} introduced a BERT-based spam detection framework, integrating classifiers such as Logistic Regression, Random Forest, K-Nearest Neighbors, and SVM. Their results, using datasets like UCI’s Spambase and the Kaggle Spam Filter Dataset, demonstrated that BERT significantly improved spam classification, achieving a precision of 97.86\% and an F1-score of 97.84\%. Meanwhile, Labonne and Moran \cite{labonne2023} explored Large Language Models (LLMs) in spam detection, developing Spam-T5, a fine-tuned version of Flan-T5. Their work showed that Spam-T5 performed exceptionally well in low-data settings, surpassing both traditional ML models and modern LLMs like BERT. Chakraborty et al. \cite{chakraborty2024sentiment} leveraged a fine-tuned BERT model with interval Type-2 fuzzy logic for sentiment classification, achieving superior performance in handling contextual variations. Similarly, Zhang et al. \cite{zhang2024hybrid} integrated BERT with large LLMs in a hybrid approach, improving sentiment intensity prediction and aspect extraction. In the domain of ethical content detection, Aziz et al. \cite{aziz2024unifying} applied BERT with multi-layered graph convolutional networks to identify sentiment triplets, highlighting the model’s capability in detecting hate speech and ethically sensitive content. These studies reinforce the adaptability of transformer-based architectures in capturing complex linguistic patterns and moral nuances, making them well-suited for ethical content classification. 

Hendrycks et al. \cite{hendrycks2021ethics} introduced the ETHICS dataset \cite{dataset} to evaluate AI models' ability to reason about morality across different ethical frameworks, including justice, virtue ethics, deontology, utilitarianism, and commonsense morality. Their findings indicate that pre-trained LLMs like BERT and RoBERTa show only partial success in making ethical decisions and often fail to handle complex moral scenarios accurately. Even advanced models like RoBERTa-large and ALBERT-xxlarge demonstrated low accuracy, particularly on adversarial test cases, highlighting their limitations in generalizing ethical principles. A key issue with the dataset is that the utilitarianism subset lacks explicit labels, requiring models to infer relative rankings rather than performing direct classification. Additionally, the study relied on standard fine-tuning techniques, but accuracy could likely improve with more extensive fine-tuning and domain-specific training. These limitations suggest that current models still struggle to integrate ethical reasoning effectively.

Pre-trained LLMs have proven highly effective in understanding complex language and context for tasks like spam detection, hate speech recognition, offensive language identification, sentiment analysis, and other forms of harmful content detection. Their adaptability and precision make them well-suited for content moderation. Building on their success, this study applies pre-trained LLMs to ethical content classification, aiming for a more reliable, context-aware, and fair moderation system.

\section{Methodology}
\label{sec:methodology}
\begin{figure}[h]
\centerline
{\includegraphics[width=\linewidth]{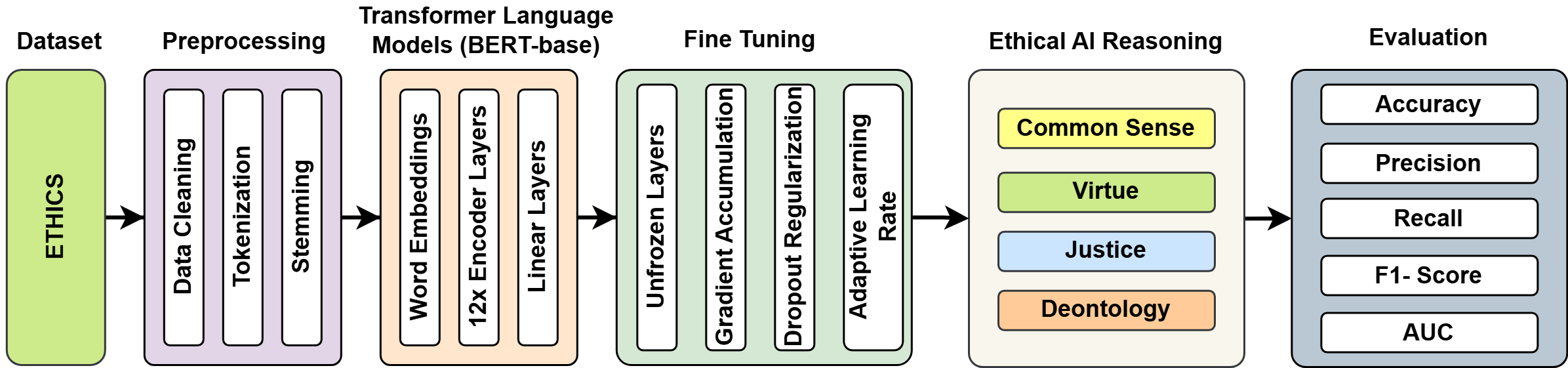}}
\caption{Overview of the methodology for ethical reasoning}
\label{fig:methodology}
\end{figure}

This section explains the methodology used in our research to analyze ethical reasoning using machine learning techniques. It includes details about the dataset, data preprocessing, and implementation of our machine learning pipeline. Additionally, we elaborate on the fine-tuning process, showcasing the innovations that adapt the pre-trained BERT model for the task of ethical reasoning analysis, as illustrated in Figure~\ref{fig:methodology}.

\subsection{Dataset}
We used the ETHICS dataset \cite{dataset}, as the foundation for this research. The ETHICS dataset is publicly available at GitHub\footnote{ \url{https://github.com/hendrycks/ethics}.}. The dataset is specifically designed to evaluate ethical reasoning in text across four key domains: Justice, Virtue, Deontology, and Commonsense. Each text sample is annotated with a corresponding ethical category label.

\begin{table}[ht!]
\caption{Data Distribution Across Splits}\label{tab:data_distribution}
\begin{tabular}{|c|c|c|c|c|}
\hline
\textbf{Split} & \textbf{\textit{Justice}} & \textbf{\textit{Virtue}} & \textbf{\textit{Deontology}} & \textbf{\textit{Commonsense}} \\
\hline
Training (80\%)       & 17432 & 22596 & 14531 & 11128 \\
\hline
Validation (20\%)     & 4359  & 5649  & 3633  & 2782  \\
\hline
Test                  & 2704  & 4975  & 3596  & 3885  \\
\hline
Hard Test             & 2052  & 4780  & 3536  & 3964  \\
\hline
\end{tabular}
\end{table}

The dataset is divided into three predefined splits: training, test, and hard test sets. To create a validation set, we further divided the original training set, allocating 80\% for training and 20\% for validation. Table~\ref{tab:data_distribution} presents the distribution of data across these splits, with the adjusted training and validation sets reflecting this allocation. The dataset’s structure and balanced representation across ethical categories ensure a reliable benchmark for training, validation, and testing. The ``Hard Test'' dataset is an adversarially filtered dataset for ethical reasoning which allows us to evaluate the model's robustness and ability to handle challenging ethical reasoning tasks.

\subsection{Adversarially filtered dataset (Hard Test)}
The Hard Test dataset is a specialized subset of the ETHICS dataset, created to evaluate AI models' ability to process complex ethical scenarios \cite{dataset}. Unlike standard test sets, it filters out straightforward cases, ensuring models rely on deeper moral reasoning rather than statistical shortcuts. The dataset was developed using an adversarial filtration process, where models (Distil-BERT and Distil-RoBERTa) were first trained on a development set and then tested on various examples \cite{hendrycks2021ethics}. The most challenging cases were those with the highest prediction errors were selected while easier examples were removed, ensuring a rigorous evaluation benchmark. This dataset is widely applicable in AI ethics research, decision-making systems, and policy-driven AI, helping assess whether models can navigate nuanced ethical dilemmas. Researchers can fine-tune AI models on broader datasets and then use the Hard Test set to measure their ability to generalize moral principles and make fairer decisions. By focusing on difficult, context-rich scenarios, the Hard Test dataset advances research in ethical AI, contributing to the development of responsible and just decision-making systems \cite{hendrycks2021ethics}.

\subsection{Data Preprocessing}
During the training of our BERT-based model, we employed an optimized data preprocessing pipeline to enhance generalization and mitigate data sparsity. These modifications were applied dynamically during training, ensuring efficient token representation, computational optimization, and contextual consistency.

\subsubsection{Text Normalization}
Text normalization improves model training by reducing vocabulary sparsity, ensuring consistency, and removing noise. It standardizes text by handling case normalization, expanding contractions, and cleaning unnecessary characters. For each input sequence \( X \), we applied a transformation function \( f : X \to X' \), where \( X' \) represents the normalized text \cite{colinraffel}. The transformation was defined as:
\begin{equation}
    X' = f(X) = g_1(X) + g_2(X) + g_3(X)
    \label{eq:norm}
\end{equation}
In Equation \ref{eq:norm}, \( g_1(X) \) handled adaptive case normalization, \( g_2(X) \) expanded contractions (e.g., \textit{can't} to \textit{cannot}), and \( g_3(X) \) removed unnecessary special characters while preserving syntactic structure. By applying these transformations during training, we reduced vocabulary sparsity and improved token consistency across training epochs \cite{normalization10447169}.

\subsubsection{Tokenization Using WordPiece}
WordPiece tokenization improves model training by handling unseen words, reducing vocabulary size, and enhancing embedding stability. It splits words into subwords based on frequency, preventing excessive fragmentation while maintaining meaningful representations. This helps models generalize better to rare and new words, making training more efficient and robust. In this process the input text was tokenized dynamically using BERT’s WordPiece algorithm \cite{tokenizationW10692355}. Each tokenized word \( w \) was decomposed into subword tokens. In Equation \ref{eq:token1}, \( V \) is BERT’s fixed vocabulary. Instead of relying on standard segmentation, we employed frequency-aware tokenization, ensuring subwords were split efficiently based on their corpus occurrence. In Equation \ref{eq:token2} \( P(T \mid w) \) denotes the probability of a subword sequence given a word. This prevented excessive fragmentation of rare words and improved embedding stability. During training, this adjustment helped the model generalize better to unseen words \cite{vaswani2017attention}.
\begin{equation}
    T_w = \{ t_1, t_2, \dots, t_n \}, \quad t_i \in V
    \label{eq:token1}
\end{equation}
\begin{equation}
    T'_w = \arg\max_T P(T \mid w)
    \label{eq:token2}
\end{equation}

\subsubsection{Truncation and Padding Optimization}
Since BERT requires a fixed sequence length \( L \), we dynamically truncated or padded input sequences during training. Padding was applied only when necessary. In Equation \ref{eq:padding}, a sequence \( S' \) is shorter than \( L \), padding tokens (\([\text{PAD}]\)) are appended. The exponent notation \( (L - |S'|) \) represents the number of padding tokens added to match the fixed length \( L \). For example, if \( S' \) has 8 tokens but \( L = 12 \), then 4 \([\text{PAD}]\) tokens are appended. To prevent overfitting due to excessive padding, we implemented batch-wise dynamic padding, which ensured that the sequence length \( L \) was adjusted based on the longest sequence in each batch. This minimized redundant \([\text{PAD}]\) tokens, leading to faster training and reduced computational overhead \cite{tunstall2022natural, sun2022skip}.
\begin{equation}
    S' = S' + [\text{PAD}]^{(L - |S'|)}
    \label{eq:padding}
\end{equation}

\subsection{Selecting a BERT-Based Cased Model}
When selecting a model for AI ethical reasoning tasks, the choice must ensure accurate interpretation and context retention. A BERT-based cased model proposed by Devlin et al. \cite{devlin2018bert}, is particularly effective due to its ability to preserve case distinctions, which are often vital in formal and ethical text analysis. This ensures that proper nouns, legal terms, and acronyms retain their intended meanings, reducing ambiguity in ethical and policy analysis \cite{rogers2020primer}. Research highlights the importance of case sensitivity in legal and ethical texts, as it helps differentiate between terms like ``Title IX'' and ``title ix'' or ``US'' and ``us,'' preventing misinterpretation. Case-sensitive models also enhance bias detection and policy evaluation by preserving textual integrity \cite{weidinger2022taxonomy}. By leveraging this approach, we improve the accuracy and reliability of our ethical assessments.

\subsection{Implementation Details}

\begin{figure}[h]
\centerline
{\includegraphics [width=\linewidth]{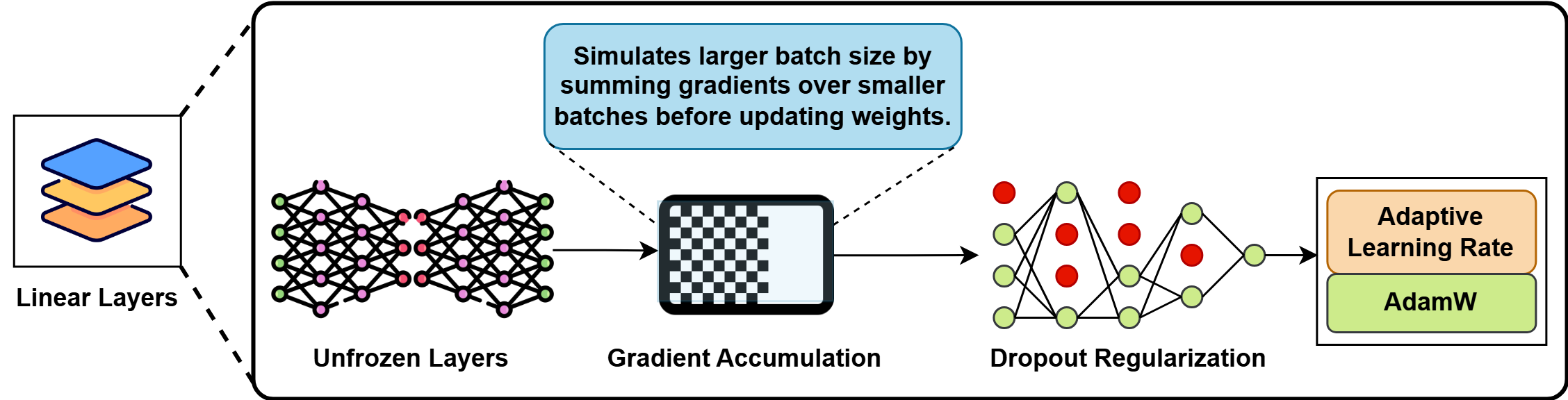}}
\caption{Fine tuning of BERT for ethical reasoning}
\label{fig:finetunning}
\end{figure}

Our implementation is centered around a fine-tuned BERT-based cased model, chosen for its strong contextual understanding and adaptability to text classification tasks. In the following, we detail the architecture, training process, and fine-tuning innovations, along with the mathematical formulations underpinning these methods illustrated in Figure \ref{fig:finetunning}. Table \ref{tab:finetuning_math} shows the customized hyperparameters and techniques employed during the fine-tuning process that ensured optimal performance on the ethical reasoning task.

\subsubsection{Model Architecture}

We used a pre-trained BERT base cased model \cite{devlin2018bert} and extended it with a classification head designed specifically for ethical reasoning tasks. This classification head consists of a fully connected layer.
\begin{equation}
    \hat{Y} = \sigma(W H_L + b)
    \label{eq:unfrezing}
\end{equation}
In Equation \ref{eq:unfrezing}, $H_L$ is the final hidden state of the transformer output. $W$ and $b$ are the weight matrix and bias vector of the classification head, and $\sigma$ is the sigmoid activation function, converting logits to probabilities for binary classification. The resulting probabilities $\hat{Y}$ represent the model’s confidence in each ethical reasoning category.

\begin{table}[ht!]
\caption{Customized Hyperparameters and Techniques for BERT Fine-Tuning}\label{tab:finetuning_math}
\begin{tabular}{|l|p{3.2cm}|p{4.3cm}|}
\hline
\textbf{Hyperparameter} & \textbf{Default} & \textbf{Modified} \\ \hline
Model Architecture & BERT based cased & BERT based cased (full unfreezing) \\ \hline
Learning Rate & 0.00002 & 0.00006 \\ \hline
Learning Rate Scheduler & None & Adaptive Learning Rate Scheduler; $\eta_t = \eta_0 \cdot \frac{1}{\sqrt{t}}$ \\ \hline
Max Sequence Length & 512 & 128 \\ \hline
Dropout Regularization & None & 0.3 \\ \hline
Gradient Accumulation & None & Accumulated over 4 mini-batches \\ \hline
Batch Size & 16 & 32 \\ \hline
Optimizer & AdamW & AdamW with Weight Decay \\ \hline
\end{tabular}
\end{table}

\subsubsection{Training Process}

The model is trained to minimize a binary cross-entropy loss $\mathcal{L}$. In Equation \ref{eq:training}, $N$ is the number of samples, $y_i$ is the true label for the $i$-th sample, and $\hat{y}_i$ is the predicted probability for the $i$-th sample. The optimizer used is AdamW with a learning rate of $\eta = 0.00006$ and applied weight decay. At each training step $t$, the parameters $\theta$ are updated as Equation \ref{eq:training}:
\begin{equation}
    \mathcal{L} = -\frac{1}{N} \sum_{i=1}^N \left[ y_i \log(\hat{y}_i) + (1-y_i) \log(1-\hat{y}_i) \right]
    \label{eq:training}
\end{equation}
\begin{equation}
    \theta^{(t+1)} = \theta^{(t)} - \frac{\eta}{\sqrt{\hat{v}_t} + \epsilon} \hat{m}_t
    \label{eq:training2}
\end{equation}
Equation \ref{eq:training2}, $\hat{m}_t$ and $\hat{v}_t$ are bias-corrected estimates of the first and second moments of gradients, and $\epsilon$ is a small constant for numerical stability.

\subsubsection{Fine-Tuning Innovations}

To maximize the model’s adaptability to the ethical reasoning task, we implemented the following innovations:

\textbf{Full Fine-Tuning:} All layers of the BERT model were unfrozen, allowing parameter adjustments across the entire network \cite{howard2018universal}. From Equation \ref{eq:finetune}, the loss gradient $\nabla_\theta \mathcal{L}$ was backpropagated through all layers \cite{devlin2018bert} where, $L$ is the number of transformer layers. Fully fine-tuning in ethical classification tasks helps the model grasp domain-specific ethical nuances, leading to more precise and fair decisions. It refines the model’s understanding beyond general pre-trained knowledge, aligning it with ethical guidelines. This approach minimizes bias, enhances reliability, and ensures responsible decision-making.
\begin{equation}
    \frac{\partial \mathcal{L}}{\partial \theta_i} = \frac{\partial \mathcal{L}}{\partial H_L} \cdot \prod_{j=i+1}^L \frac{\partial H_j}{\partial H_{j-1}} \cdot \frac{\partial H_i}{\partial \theta_i}, \quad i = 1, 2, \dots, L
    \label{eq:finetune}
\end{equation}

\textbf{Gradient Accumulation:} To address memory constraints, gradient accumulation was employed. Gradients $g^{(b)}$ for each mini-batch $b$ were accumulated over $N_\text{acc}$ steps \cite{reddi2019convergence}. Equation \ref{eq:gradient1}, gradients \( \nabla \mathcal{L}^{(b)} \) from each mini-batch \( b \) are summed over \( N_{\text{acc}} \) steps. This method allows training with small mini-batches while effectively simulating a larger batch size. Equation \ref{eq:gradient2} updates the model parameters \( \theta \) after accumulating gradients over multiple steps. The learning rate \( \eta \) scales the average accumulated gradient, ensuring stable optimization. It ensures better representation of ethical considerations in data while maintaining computational feasibility. The approach helps to mitigate biases and enhances fairness by enabling effective learning from smaller yet diverse datasets.
\begin{equation}
    g_\text{acc} = \sum_{b=1}^{N_\text{acc}} \nabla_\theta \mathcal{L}^{(b)}    
    \label{eq:gradient1}
\end{equation}
\begin{equation}
    \theta^{(t+1)} = \theta^{(t)} - \eta \frac{g_\text{acc}}{N_\text{acc}}
    \label{eq:gradient2}
\end{equation}

\textbf{Adaptive Learning Rate:} An adaptive learning rate schedule was used, reducing the learning rate as training progressed \cite{vaswani2017attention}. In Equation \ref{eq:adaptive_learning}, $\eta_0$ is the initial learning rate, and $t$ is the training step. Applying an adaptive learning rate in model training dynamically adjusts the step size based on gradient variations, improving convergence speed and stability. This technique helps prevent overshooting in high-gradient regions while accelerating learning in flatter areas, leading to more efficient optimization. In ethical classification tasks, adaptive learning rates enhance fairness and robustness by ensuring balanced learning across diverse and sensitive data distributions.
\begin{equation}
    \eta_t = \eta_0 \cdot \frac{1}{\sqrt{t}}    
    \label{eq:adaptive_learning}
\end{equation}

\subsubsection{Regularization and Robustness}

Dropout regularization was applied in the classification head to mitigate overfitting. During training \cite{wager2013dropout}, activations $H_\text{task}$ in the classification head were stochastically zeroed out. In Equation \ref{eq:regularization}, $D \sim \text{Bernoulli}(1-p)$ is a dropout mask, $\odot$ represents element-wise multiplication, and $p$ is the dropout rate. At inference time, activations were scaled by $(1-p)$ to maintain consistent output expectations shown in Equation \ref{eq:regularization2}. Regularization techniques, dropout, and batch normalization, help prevent overfitting by constraining model complexity. These methods ensure that the model generalizes well to unseen ethical scenarios, reducing biases and improving fairness. In ethical classification tasks, regularization enhances robustness by making the model resilient to noisy or imbalanced data, leading to more reliable and ethically sound decisions.

\begin{equation}
    H'_\text{task} = D \odot H_\text{task}
    \label{eq:regularization}
\end{equation}
\begin{equation}
    H_\text{task}^\text{inference} = (1-p) \cdot H_\text{task}
    \label{eq:regularization2}    
\end{equation}

\subsection{Evaluation Matrix}
The model’s performance was evaluated using accuracy, precision, recall, F1-score \cite{powers2020evaluation}, and AUC \cite{namdar2021modified} ensuring robust validation of ethical reasoning capabilities.

\section{Results and Discussion}

This section presents the evaluation of the fine-tuned BERT model on the Test Split and Hard Test Split datasets, along with a comparison to existing models such as RoBERTa-large and ALBERT-xxlarge. The results demonstrate the impact of our innovative fine-tuning and preprocessing techniques in delivering strong performance, particularly in specific ethical reasoning domains.

\subsection{Performance on Test Split}

\begin{figure}[h]
    \centering
    \begin{subfigure}[b]{0.45\linewidth}
        \centering
        \includegraphics[width=\linewidth]{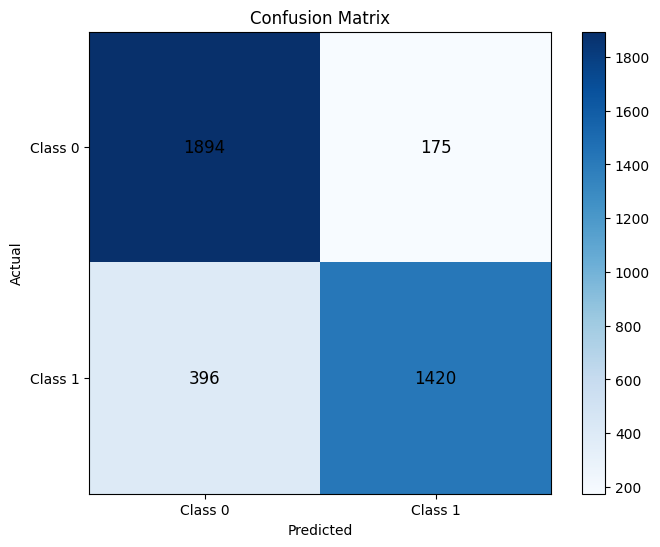}
        \caption{Common Sense}
        \label{fig:test_common}
    \end{subfigure}
    \hfill
    \begin{subfigure}[b]{0.45\linewidth}
        \centering
        \includegraphics[width=\linewidth]{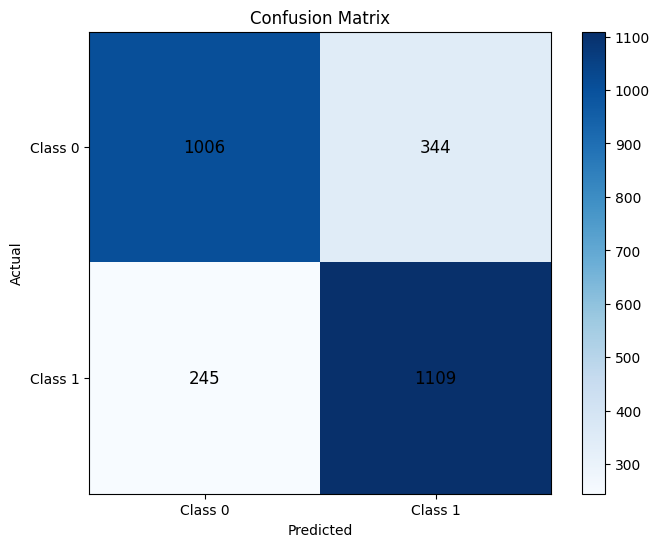}
        \caption{Justice}
        \label{fig:test_justice}
    \end{subfigure}
    \hfill
    \begin{subfigure}[b]{0.45\linewidth}
        \centering
        \includegraphics[width=\linewidth]{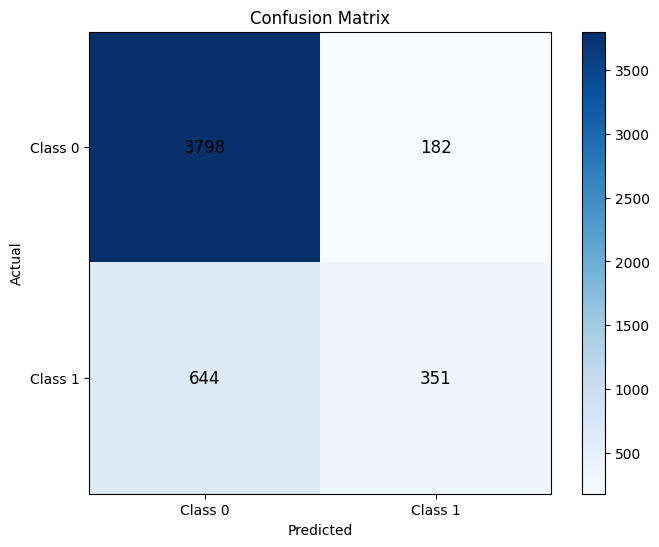}
        \caption{Virtue}
        \label{fig:test_virtue}
    \end{subfigure}
    \hfill
    \begin{subfigure}[b]{0.45\linewidth}
        \centering
        \includegraphics[width=\linewidth]{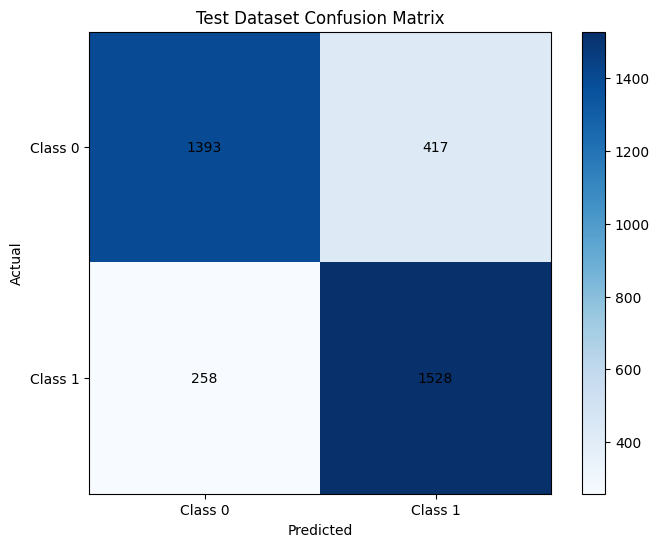}
        \caption{Deontology}
        \label{fig:test_deontology}
    \end{subfigure}
    \caption{Confusion matrix of Test dataset.}
    \label{fig:testconfusionmatrix}
\end{figure}

Table~\ref{tab:result_test} shows the performance of the proposed BERT model on the Test Split. The model achieved high Accuracy in Commonsense, Justice, Virtue domains and Deontology, reaching 86.46\%, 78.22\%, 83.40\%, and 81.23\% respectively. These results highlight the model's ability to effectively adapt to the task in these domains. The AUC values for domains—90.78, 87.36, 88.78, 89.93—further affirm the model’s capability to separate positive and negative classes accurately. The confusion matrices for the Test dataset are presented in Figure \ref{fig:testconfusionmatrix}. 

The subfigures \ref{fig:test_common}, \ref{fig:test_justice}, \ref{fig:test_virtue}, and \ref{fig:test_deontology} represent the results of the Common Sense, Justice, Virtue, and Deontology frameworks, respectively, demonstrating variations in the performance of the model in different paradigms of ethical reasoning.

\begin{table}[ht!]
\caption{Performance of BERT over Test Split Data}\label{tab:result_test}
\begin{tabular}{|c|c|c|c|c|c|}
\hline
\textbf{Sub-set} & \textbf{Accuracy} & \textbf{Precision} & \textbf{Recall} & \textbf{F1-score} & \textbf{AUC} \\
\hline
Commonsense & 86.46 & 86.00 & 85 & 85 & 90.78 \\
Justice & 78.22 & 78 & 78 & 78 & 87.36 \\
Virtue & 83.40 & 82 & 83 & 81 & 88.78 \\
Deontology & 81.23 & 78.56 & 85.55 & 81.00 & 89.93 \\
\hline
\end{tabular}
\end{table}

\subsection{Performance on Hard Test Split}

The Hard Test Split, designed to assess robustness on more challenging examples, revealed further insights into the model’s strengths and weaknesses, as summarized in Table~\ref{tab:result_hardtest}. The Accuracy in Commonsense domain dropped to 50.00\%, indicating difficulty in handling complex or ambiguous cases. This suggests a need for additional strategies to improve generalization in these domains.

\begin{figure*}[htbp]
    \centering
    \begin{subfigure}[b]{0.45\linewidth}
        \centering
        \includegraphics[width=\linewidth]{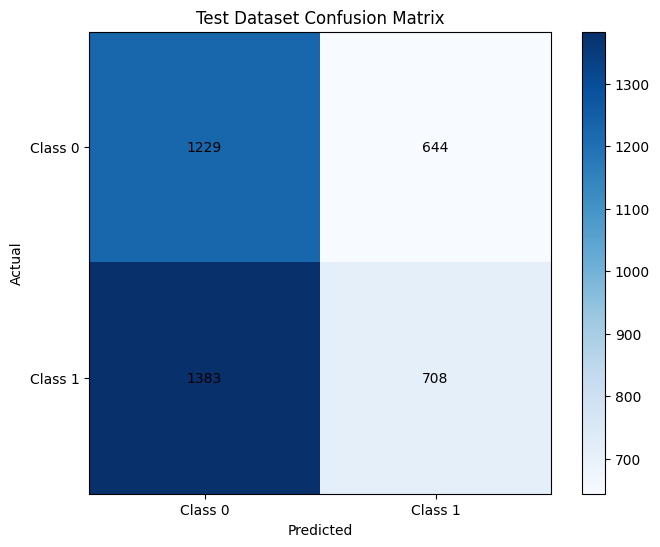}
        \caption{Common Sense}
        \label{fig:hardtest_common}
    \end{subfigure}
    \hfill
    \begin{subfigure}[b]{0.45\linewidth}
        \centering
        \includegraphics[width=\linewidth]{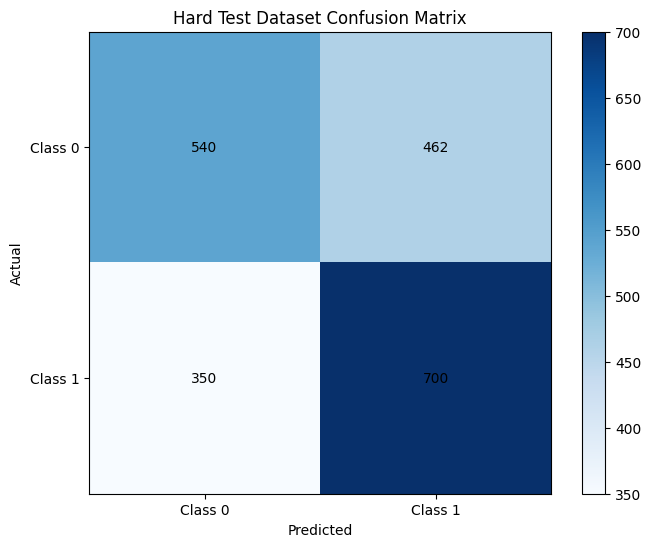}
        \caption{Justice}
        \label{fig:hardtest_justice}
    \end{subfigure}
    \hfill
    \begin{subfigure}[b]{0.45\linewidth}
        \centering
        \includegraphics[width=\linewidth]{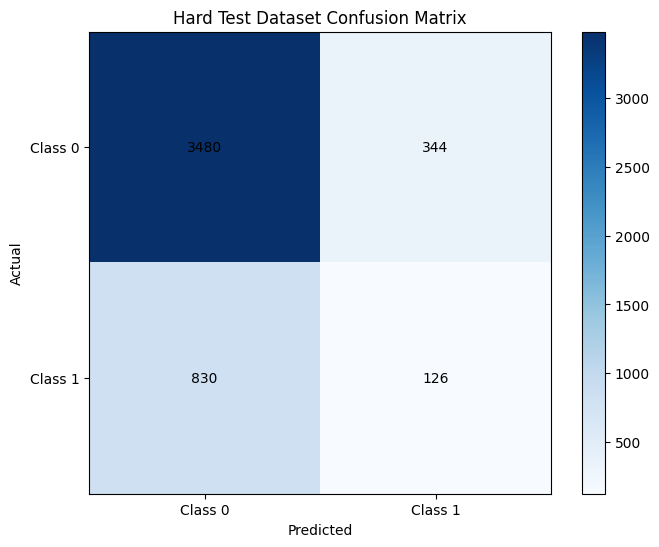}
        \caption{Virtue}
        \label{fig:hardtest_virtue}
    \end{subfigure}
    \hfill
    \begin{subfigure}[b]{0.45\linewidth}
        \centering
        \includegraphics[width=\linewidth]{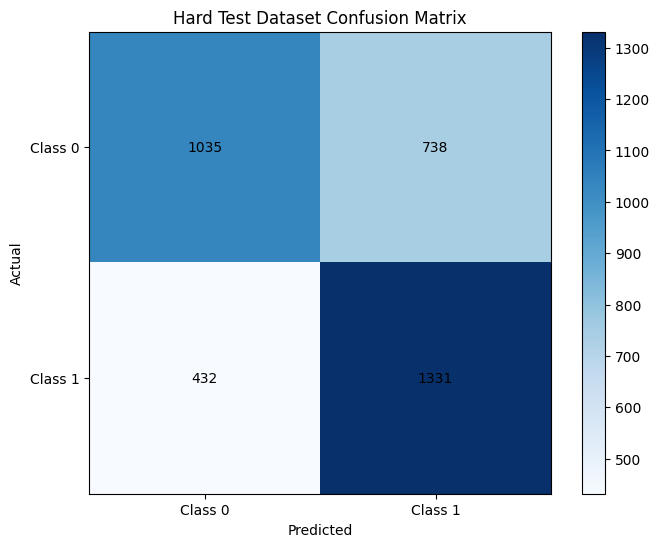}
        \caption{Deontology}
        \label{fig:hardtest_deontology}
    \end{subfigure}
    \caption{Confusion matrix of Hard Test dataset.}
    \label{fig:hardtestconfusionmatrix}
\end{figure*}

In contrast, the model maintained stronger performance in the Deontology and Virtue domains, achieving 66.91\% and 75.44\% Accuracy, respectively. The AUC values of 73.35 for Deontology and 80.24 for Virtue further underscore the model’s resilience. These results highlight the role of our preprocessing and training techniques in ensuring that the model retains its effectiveness even with harder data. The confusion matrices for the Hard Test dataset are presented in Figure \ref{fig:hardtestconfusionmatrix}. 

\begin{figure}[htpb]
    \centering
    \begin{subfigure}[b]{0.88\linewidth}
        \centering
        \includegraphics[width=\linewidth]{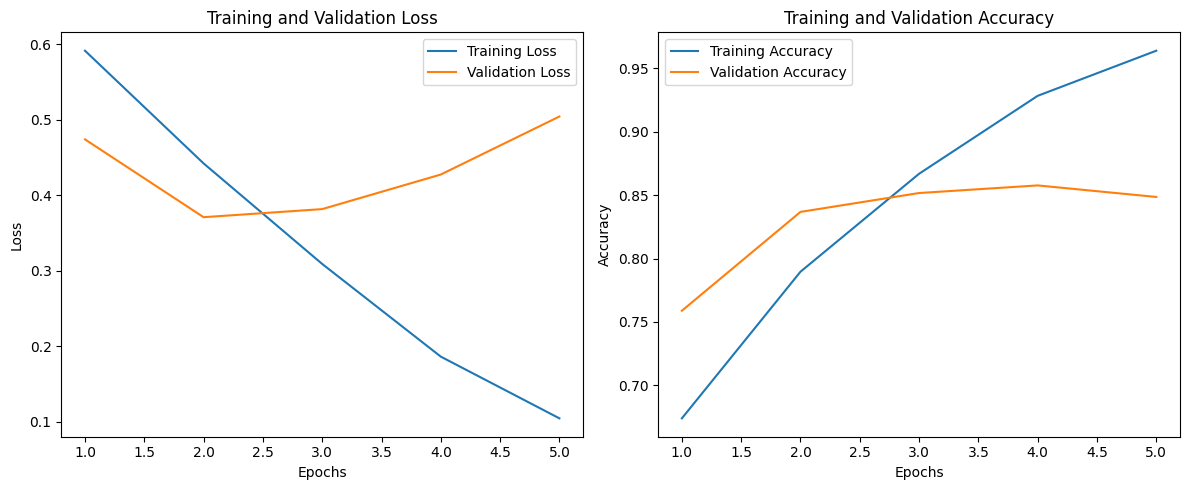}
        \caption{Common Sense}
        \label{fig:curve_common}
    \end{subfigure}
    \hfill
    \begin{subfigure}[b]{0.88\linewidth}
        \centering
        \includegraphics[width=\linewidth]{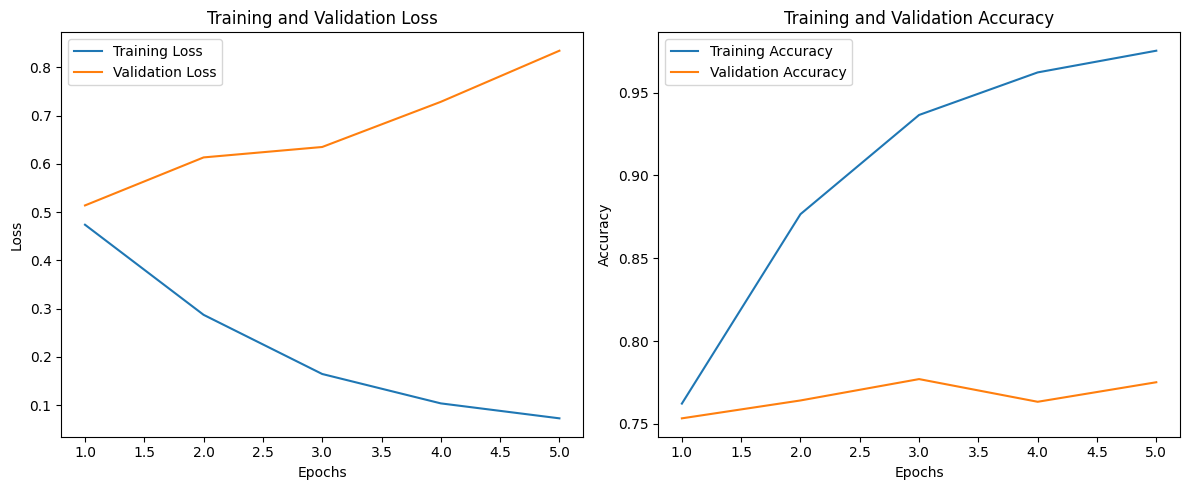}
        \caption{Justice}
        \label{fig:curve_justice}
    \end{subfigure}
    \hfill
    \begin{subfigure}[b]{0.88\linewidth}
        \centering
        \includegraphics[width=\linewidth]{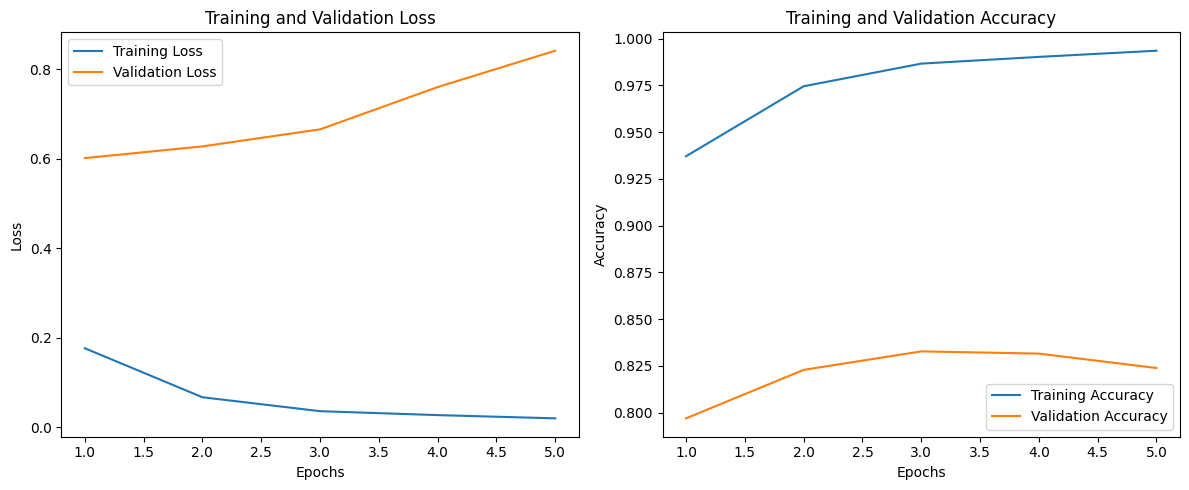}
        \caption{Virtue}
        \label{fig:curve_virtue}
    \end{subfigure}
    \hfill
    \begin{subfigure}[b]{0.88\linewidth}
        \centering
        \includegraphics[width=\linewidth]{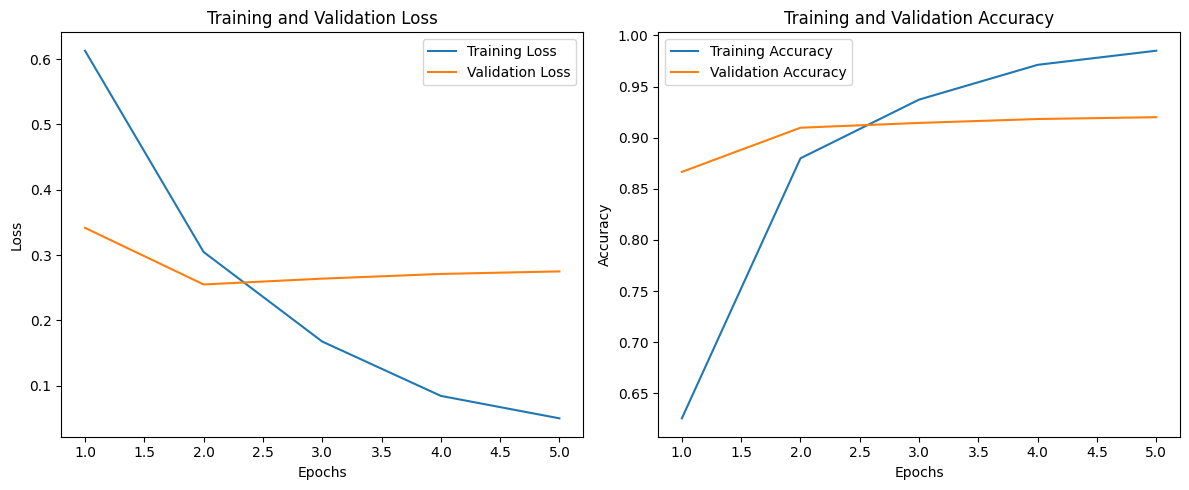}
        \caption{Deontology}
        \label{fig:curve_deontology}
    \end{subfigure}
    \caption{Training vs validation loss and accuracy curves on training dataset.}
    \label{fig:curve}
\end{figure}

The subfigures \ref{fig:hardtest_common}, \ref{fig:hardtest_justice}, \ref{fig:hardtest_virtue}, and \ref{fig:hardtest_deontology} correspond to the Common Sense, Justice, Virtue, and Deontology frameworks, respectively, highlighting differences in model performance across ethical reasoning approaches. The Figure \ref{fig:curve} illustrate model performance over five epochs, highlighting key trends. Training loss consistently decreases, while accuracy improves, indicating effective learning. Some models maintain stable validation accuracy, suggesting good generalization. In some cases, training and validation loss patterns differ, which may indicate areas for refinement. Adjustments like regularization could improve performance. Overall, the models demonstrate effective learning, with some showing stronger generalization.

\begin{table}[ht!]
\caption{Performance of BERT over Hard-Test Split Data}\label{tab:result_hardtest}
\begin{tabular}{|c|c|c|c|c|c|}
\hline
\textbf{Sub-set} & \textbf{Accuracy} & \textbf{Precision} & \textbf{Recall} & \textbf{F1-score} & \textbf{AUC} \\
\hline
Commonsense & 50.00 & 50.00 & 49 & 48 & 50.26 \\
Justice & 60.40 & 60.24 & 66.67 & 60 & 65.39 \\
Virtue & 75.44 & 70 & 75 & 72 & 80.24 \\
Deontology & 66.91 & 64.33 & 75.50 & 67.00 & 73.35 \\
\hline
\end{tabular}
\end{table}

\subsection{Comparison with Existing Models}

Tables~\ref{tab:comparison_test} and \ref{tab:comparison_hardtest} compare the proposed BERT model's performance with other models like RoBERTa-large and ALBERT-xxlarge. On the Test Split, our model achieved an average Accuracy of 82.32\%, significantly outperforming 
 where the baseline BERT-base got 46.1\% and RoBERTa-large 65.1\% while nearly beating ALBERT-xxlarge 68.3\%. Notably, the proposed Ethic-BERT excelled in Justice, Virtue and Deontology domains, achieving 78.22\%, 83.40\% and 81.23\% Accuracy, respectively.

\begin{table}[]
\caption{Comparison of Test Results with Existing Models}\label{tab:comparison_test}
{\begin{tabular}{@{}ccccccc@{}} \toprule
\textbf{Source} & \textbf{Model} & \textbf{\makecell{Common\\Sense}} & \textbf{Justice} & \textbf{Virtue} & \textbf{Deontology} & \textbf{Average} \\ \midrule
\multirow{3}{*}{Hendrycks et al. \cite{hendrycks2021ethics}} 
& BERT-base       & 86.5  & 26.0  & 33.1  & 38.8  & 46.1 \\
& RoBERTa-large   & 90.4  & 56.7  & 53.0  & 60.3  & 65.1 \\
& ALBERT-xxlarge  & 85.1  & 59.9  & 64.1  & 64.1  & 68.3 \\ 
\textbf{Proposed} & \textbf{Ethic-BERT} & \textbf{86.46} & \textbf{78.22} & \textbf{83.40}  & \textbf{81.23} & \textbf{82.32} \\ \bottomrule
\end{tabular}}
\end{table}

\begin{table}[]
\caption{Comparison of Hard Test Results with Existing Models}\label{tab:comparison_hardtest}
{\begin{tabular}{@{}ccccccc@{}} \toprule
\textbf{Source} & \textbf{Model} & \textbf{\makecell{Common\\Sense}} & \textbf{Justice} & \textbf{Virtue} & \textbf{Deontology} & \textbf{Average} \\ \midrule
\multirow{3}{*}{Hendrycks et al. \cite{hendrycks2021ethics}} 
& BERT-base       & 48.7  & 7.6   & 8.6   & 10.3  & 16.8 \\
& RoBERTa-large   & 63.4  & 38.0  & 25.5  & 30.8  & 39.42 \\
& ALBERT-xxlarge  & 59.0  & 38.2  & 37.8  & 37.2  & 43.05 \\ 
\textbf{Proposed} & \textbf{Ethic-BERT} & \textbf{50.00} & \textbf{60.40} & \textbf{75.44} & \textbf{66.91} & \textbf{63.18} \\ \bottomrule
\end{tabular}}
\end{table}

On the Hard Test Split, the proposed Ethic-BERT achieved an average Accuracy of 63.18\%, surpassing BERT-base 16.8\%, RoBERTa-large 39.42\%, and ALBERT-xxlarge 43.05\%. The model’s performance in Justice, Virtue and Deontology domains remained robust, with Accuracy values of 60.40\%, 75.44\% and 66.91\%. These improvements suggest that our approach enables the model to perform consistently better across harder examples compared to other models.

\subsection{Contributions of Fine-Tuning and Preprocessing}

The superior results of the proposed model are attributed to the combination of comprehensive fine-tuning and a robust preprocessing pipeline. Unlike traditional approaches that partially freeze pre-trained layers, our method unfreezes all layers of the BERT model, allowing it to fully adapt its representations to the specific nuances of ethical reasoning. This approach ensures that the model effectively incorporates both pre-trained knowledge and task-specific patterns.

The preprocessing pipeline was another critical factor in achieving these results. By employing advanced tokenization, consistent input formatting through truncation and padding, the pipeline enhanced the quality and diversity of the training data. Additionally, the use of gradient accumulation allowed for stable training even with limited resources, optimizing learning efficiency. The balanced data splits further ensured that all ethical reasoning domains were well-represented during training, contributing to strong results in Justice and Virtue categories.

\section{Challenges and Future Research Directions}

Despite the promising results, challenges remain in the deontology and common sense domains, especially in the hard-test split. The low accuracy and AUC in these domains indicate the need for additional strategies, such as augmenting the dataset with richer context or leveraging external knowledge sources. Incorporating domain-specific pretraining or multitask learning approaches may also help the model capture the unique characteristics of these categories. Although, the proposed Ethic-BERT model demonstrates substantial advancements in ethical reasoning classification, achieving state-of-the-art performance in several domains, it can further be improved which is outlined as the future research directon in this topic. The effectiveness of fine-tuning strategies highlights the potential for further improvements in AI systems designed for complex reasoning tasks.

\section{Conclusion}
In this study, we fine-tuned a BERT-based model to classify ethical reasoning across four domains: Common sense, justice, virtue, and deontology. The model demonstrated strong performance, particularly in Justice and Virtue reasoning, where it surpassed existing models. In the Hard Test Split, the model showcased its robustness, achieving improvements over baseline approaches in more challenging scenarios. Our approach introduced key innovations, including comprehensive fine-tuning by unfreezing all layers of the BERT model, implementing gradient accumulation, and utilizing advanced tokenization and data augmentation. These techniques allowed the model to effectively combine its pretrained knowledge with task-specific adaptations, resulting in superior performance. Despite these successes, challenges persist in the Commonsense and Deontology domains, especially on the Hard Test Split. Addressing these gaps could involve enriching the training data with more contextually diverse examples, incorporating external knowledge sources, or adopting domain-specific pretraining strategies. In general, this work highlights the potential of fine-tuned transformer models to tackle complex reasoning tasks in AI ethics. The findings underscore the importance of thoughtful preprocessing and training techniques in improving the robustness and generalization of the model.

\section*{Acknowledgments}
The authors would like to express their sincere gratitude to the Ubiquitous, Cloud, and Human-Computer Interaction (UCH) Research Group, Department of Computer Science, American International University-Bangladesh for supporting this research.

\bibliography{sn-bibliography}

\end{document}